\documentclass[prd,a4paper,tightenlines,eqsecnum,nofootinbib,preprint,floatfix,showpacs]{revtex4}
\usepackage{epsfig}
\usepackage{graphicx}
\usepackage{bm}

\def\Tr{\text{Tr}\,}
\def\tr{\text{tr}\,}
\def\Eq#1{Eq.~(\ref{#1})}

\def\<{\langle}
\def\>{\rangle}
\def\bn{{\Boldmath n}}
\newcommand{\text}{\rm}
\def\>{{\rangle}}
\def\<{{\langle}}

\def\bn{{\bm{n}}}
\def\bN{{\bm{N}}}

\def\muhat{\hat{\bm{\mu}}}
\def\nuhat{\hat{\bm{\nu}}}

\def\bmm{{\bm{m}}}
\def\bk{{\bm{k}}}

\def\Tr{{\rm Tr}\,}
\def\tr{{\text tr}\,}

\def\Eq#1{Eq.~(\ref{#1})}

\begin{document}

\vspace*{1.0in}

\title{The critical region of strong-coupling lattice
QCD  \\ in different large-$N$ limits \vspace*{0.25in}}
\author{Barak Bringoltz\\
\vspace*{0.25in}}

\affiliation{Rudolf Peierls Centre for Theoretical Physics,
University  of  Oxford,\\
1 Keble Road, Oxford, OX1 3NP, UK \\
\vspace*{0.6in}}

\begin{abstract}
We study the critical behavior at nonzero temperature
phase transitions of an effective Hamiltonian derived from lattice QCD in the strong-coupling expansion. Following studies of
related quantum spin systems that have a similar Hamiltonian, we show that for large $N_c$ and fixed
$g^2N_c$, mean field scaling is not expected, and that the critical
region has a finite width at $N_c=\infty$. A different behavior
rises for $N_f\to \infty$ and fixed $N_c$ and $g^2/N_f$, which we
study in two spatial dimensions and for $N_c=1$. We find that the
width of the critical region is suppressed by $1/N_f^p$ with
$p=1/2$, and argue that a generalization to $N_c>1$ and to three
dimensions will change this only in detail (e.g. the value of
$p>0$), but not in principle. We conclude by stating under what
conditions this suppression is expected, and remark on possible
realizations of this phenomenon in lattice gauge theories in the
continuum.

\end{abstract}

\maketitle

\section{Introduction and summary of results}
\label{sec:intro}

Large-$N$ expansions are useful to approach nonperturbative dynamics
of field theories and are often used to study their phase transitions at
nonzero temperature $T$. Here we focus on the dependence of the
critical region of strongly coupled lattice QCD on the number of colors, $N_c$, and
flavors, $N_f$. This region of
temperatures is where mean field (MF) theory fails, and
fluctuations, that cannot be ignored, drive the system to a
nontrivial fixed point.

What motivates us are the studies of the suppression of this region
in the large-$N$ Gross-Neveu and Yukawa models
\cite{Rosenstein,Kogut1,Kogut2}. In particular, the authors in
\cite{Kogut2}, suggest that the same phenomenon occurs in QCD at
large-$N_c$, and discuss its physical implications. In apparent
contrast to that, in the letter \cite{Chandrasekharan}, which
presents numerical simulations of strongly-coupled lattice QCD with
staggered fermions and large $N_c\le 48$, the authors claim to find
that MF fails at large-$N_c$, and as evidence show that the critical
region does not shrink with $N_c$ in their simulations.

To understand this apparent contradiction and when to expect a suppression
of the critical region in general, we study an effective Hamiltonian which is derived in
second order of the strong-coupling expansion from the lattice QCD Hamiltonian, and that describes the
low energy effective excitations of mesons.
We work in $d$ spatial dimensions and
with $N_c$ colors and $N_f$ flavors of naive fermions. This gives us
flexibility to explore a large parameter space, and enables us to
show that which large-$N$ makes a system trivial depends on which
$N$ is large. Following studies of similar Hamiltonians in the
condensed matter literature \cite{MA1,AA,RS,MA2,RS1}, we explore the
critical behavior of this effective Hamiltonian in three different large-$N$ limits: (I) the large-$N_c$ limit, where
the `t Hooft coupling $g^2N_c$ is kept fixed. (II) the combined limit of large-$N_c$ {\em
and} large-$N_f$, where a l\'a Veneziano, we fix both $N_c/N_f$ and $g^2N_c$, and
(III) the limit of large $N_f$ but with fixed $N_c$ and $g^2/N_f$.
We summarize the results of these studies in the next paragraphs.

In the large-$N_c$ limit we show that the largeness of $N_c$ serves to
suppress quantum fluctuations, and leads to a classical Hamiltonian.
This Hamiltonian is a generalized Heisenberg antiferromagnet, whose
ordered ground state corresponds to the chiral broken phase. A MF
ansatz in terms of the `spins' is exact only at $T=0$, and is not
adequate to study the critical behavior near $T_c$. In particular,
the critical exponents are determined by $N_f$ and $d$, and are not
expected to belong to the Gaussian fixed point. One can, {\em as a
first step}, analyze the transition in MF, but by definition, this
analysis is bound to fail within the critical region, which we show
to have a nonzero width at $N_c=\infty$. Unfortunately, since at $N_c=\infty$ the temperature $T_c$ turns out to be
infinite, then it is difficult to learn about the behavior of transitions in planar QCD
 from the `t Hooft limit of our effective Hamiltonian, and we therefore use these results
only to understand \cite{Chandrasekharan}, but discuss how they should be improved.

In the combined limit of large $N_c$ and $N_f$, the situation is
different, and one can solve for a MF ansatz that is exact at
$N_f,N_c\to \infty$ for all $T$ ($T_c$ is finite is this limit). Reviewing known results, we show that
this solution yields a correlation length that diverges with a MF exponent only for $d>4$,
and otherwise for $2<d\le4$. Extending this, we also show that the
width of the critical region depends on $N_f$, and $N_c$ only
through their ratio $N_c/N_f$, and is nonzero.

 In the limit of large-$N_f$ and fixed $N_c$, a MF ground state
 is again exact for all $T$, but leaves chiral symmetry intact even
 at $T=0$. In $d=2$, a global minimum of the system is the ``spin-Peierls'' state, which breaks
lattice translations and rotations \cite{MA2,RS}. A corresponding
analysis in $d=3$ includes an extensive search in the space of all
possible ansatze for the global ground state, and is out of the
scope of this work. Since this limit has a restricted relevance for
QCD (for example, asymptotic freedom is lost here) but is crucial
for our purpose (see below), we proceed to study the restoration of
these lattice symmetries at finite $T$ for the $d=2$ system, and for
simplicity also fix $N_c=1$. We believe that a generalization to
$d=3$, and $N_c > 1$ will change the results only in detail but not
in principle, and find no point to do so. (For example, see the
$N_c>1$ generalization at $T=0$, and $d=2$ in \cite{RS})

The lattice symmetries are restored in a second order transition at the finite
$T_c$, whose critical behavior is closely related to the behavior
seen in \cite{Rosenstein,Kogut2,Pelissetto1,Pelissetto2} for the
systems studied there. We show that taking $N_f\to \infty$ before
$t\equiv |T-T_c|/T_c\to 0$, makes MF exponents exact. Switching the
order of the limits, we find that the Landau-Ginzburg-Wilson (LGW)
action for the transition describes scalar fields coupled through
$O(1/N_f)$ interactions. As emphasized in \cite{Kogut2,Pelissetto1},
 this suggests a crossover behavior, which we find to occur when $t\sim
1/\sqrt{N_f}$.

Combining the results of \cite{Kogut2,Pelissetto1,Pelissetto2}, and
of our study here, we emphasize that the critical region in
large-$N$ phase transitions is suppressed only when the LGW
effective action for the transition has an overall factor of
$N^\alpha, \alpha>0$. This suppresses thermal fluctuations in the
order parameter, and can make MF scaling exact. Our message here is
that this does not happen in all large-$N$ treatments, and in
particular does not happen for the chiral phase transitions of
 the effective Hamiltonian that we study here.

 We begin this report in
Section~\ref{sec:Heff} where we introduce the effective Hamiltonian
of lattice QCD in the strong-coupling limit. We then move to discuss
large $N_c$ and fixed $N_f$ in Section~\ref{sec:tHooft}, large $N_c$
and large $N_f$ in Section~\ref{sec:Veneziano}, and the large $N_f$ with
fixed $N_c$ in Section~\ref{sec:largeNf}. We conclude in
Section~\ref{sec:summary}, and make proposals for future research in
Section~\ref{sec:future}.

\section{The strong-coupling limit: the effective Hamiltonian}
\label{sec:Heff}

Strong coupling expansions were used in lattice gauge theory since its early days. In particular, they were performed
for the gauge group $SU(N_c)$ as well as for $SU(3)$. For example, in the pure $SU(N_c)$ gauge theory, expansions were made for
the string tension and the for free energy as well
as for the beta function \cite{strong_pure_gauge}. In the case of QCD with $N_c$ colors, and various types of fermions,
calculations were made for the effective low energy action of hadrons and for mesons masses \cite{Smit,strong_QCD}. The
result of these treatments (which is most relevant to the discussions is Sections~\ref{sec:tHooft}-\ref{sec:Veneziano}) is that for large values of $N_c$, the natural
expansion parameter in the strong coupling series is the inverse `t Hooft coupling $1/(g^2N_c)$ rather than $1/g^2$
 (see also \cite{BG,tHooft}).
 More precisely, in these expansions one fixes $N_c$ and take $g^2N_c\gg1$ to find that
 quantities which are $O(N_c^0)$ at large-$N_c$, such as $m_{\text meson}$,
$m_{\text baryon}/N_c$, $f^2_{\pi}/N_c$, etc, can be written as a power series in
$(g^2N_c)^{-1}$ with coefficients that depend on $N_c$ and that
become finite at $N_c=\infty$.

The result of this procedure may be different from taking the
opposite order of the limits (first $N_c\to \infty$ and only then
$g^2N_c\to \infty$). As discussed in \cite{BG,GW}, one may find
that these limits commute only as long as $(g^2N_c)_0 < g^2N_c <
\infty$. Indeed for the
two-dimensional pure gauge theory with the Wilson action one finds
$(g^2N_c)_0=2$ \cite{GW}. For higher dimensions the situation is more
involved but there still is a finite range of $g^2N_c$ in which the
theory is in a strong coupling phase (For example see early
analytical results in \cite{GS} and recent numerical results in
\cite{bulk_numerical}). Since fermions are quenched in the `t Hooft limit, we take these
results to apply for QCD (with a fixed nonzero mass) as well, but we are not aware of any analogous results in the
Veneziano limit of fixed $N_f/N_c$.

\begin{figure}[htb]
\begin{center}
        \epsfig{width=3cm,file=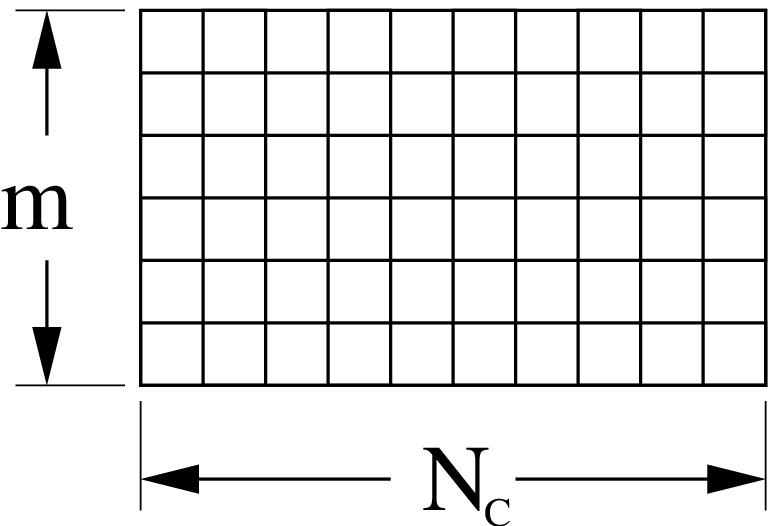} \caption{The representation
          of $U(4N_f)$ carried by $Q^{\eta}$. $m$ is related to the
          baryon number at the
site according to $m=B+2N_f$, with $|B|\leq 2N_f$.}\label{fig:Young}
\end{center}
\end{figure}

In this paper we follow the former approach (which is also the approach of Euclidean treatments such as
\cite{Chandrasekharan}): we fix $N_c$ and use an expansion in $(g^2N_c)^{-1}\ll 1$.
The result of this expansion in the Hamiltonian formalism is that at second
order one obtains the following effective Hamiltonian,
\begin{equation}
H=\frac{J}{N_c} \sum_{\bn}\sum_{\mu}\sum_{\eta} Q^\eta_\bn
Q^\eta_{\bn+\muhat}. \label{eq:H_AFM}
\end{equation}
which is derived for naive fermions in
\cite{Smit,Ben}. Here $\bn$ denotes lattice sites,
$\muhat=\hat{x}_1,\dots,\hat{x}_d$ denotes the lattice directions,
and the coupling $J$ scales with the `t Hooft coupling like $(g^2N_c)^{-1}$ \cite{Smit,thesis}. The operators $Q^{\eta}$ generate the $U(4N_f)$
algebra
\begin{equation}
\left[ Q^\eta_\bn , Q^\rho_\bmm \right] = if^{\eta \rho \sigma} Q^\sigma_\bn \delta_{\bn \bmm}, \label{eq:commutation}
\end{equation}
and are defined as
\begin{equation}
Q^\eta_\bn=\sum_{\alpha \beta,a}\psi^{\dag a \alpha}_{\bn} M^\eta_{\alpha \beta} \psi^{a \beta}_{\bn}. \label{eq:Q}
\end{equation}
Here the fermion indices $\alpha$ and $\beta$ are Dirac-flavor indices that range from $1$
to $4N_f$, while $a$ is a color index that ranges from $1$ to
$N_c$. The matrices
 $M^\eta$ are the generators of $U(4N_f)$ in the fundamental
 representation, and obey
\begin{equation}
\tr M^\eta M^\rho = \frac12 \delta_{\eta \rho}, \qquad \qquad
\sum_{\eta=1}^{16N_f^2} M^\eta_{\alpha\beta}
M^\eta_{\gamma\delta}=\frac12
\delta_{\alpha\delta}\delta_{\beta\gamma}. \label{eq:M_eta}
\end{equation}
The Hilbert space on site $\bn$ is an irreducible representation of
$U(4N_f)$ that corresponds to
a rectangular Young tableau with $N_c$ columns, and $m$ rows, see Fig.~\ref{fig:Young}.
 The corresponding baryon number on
such a site is $B=m-2N_f$. In this work we restrict to
configurations
 with zero average baryon number, and in particular examine
configurations with baryon number $B$ on the even sites and
 $-B$ on the odd sites. This means that we put conjugate $U(4N_f)$
 representations on adjacent sites, represented by Young tableaux with
 $m$ rows on the even sites, and with $(N-m)$ on the odd sites.
Following \cite{AA,RS,MA1} we take $m=2N_f$ in
Sections~\ref{sec:tHooft}, and \ref{sec:largeNf}, and $m=1$ in
Section~\ref{sec:Veneziano}. The reason is that these cases are
the simplest to analyze.

The Hamiltonian \Eq{eq:H_AFM} is a generalized Heisenberg quantum
antiferromagnet, and its N\'eel ordered state spontaneously breaks
the global $U(4N_f)$ symmetry. Adding explicit symmetry breaking
terms can reduce $U(4N_f)$ to its subgroup $SU(N_f)_L\times
SU(N_f)_R\times U(1)_B$ (e.g. see \cite{Ben,NNN}), and as a result,
the N\'eel phase corresponds to the spontaneous breakdown of chiral
symmetry.\footnote{See, however, the remark at the top of
Section~\ref{sec:Veneziano}.}

Note that \Eq{eq:H_AFM} is a result of Rayleigh-Schrodinger degenerate perturbation theory
applied to split the degeneracy of the
 zero gauge-flux sector of the strong-coupling lattice QCD spectrum.
Excitations of strings with a nonzero gauge flux cost an energy that
scales like $\sim g^2N_c \gg1$ and are neglected here. As a result
the chiral phase transition of \Eq{eq:H_AFM} will not involve a
condensation of Polyakov lines of the type discussed in
\cite{Polyakov}. Adding the effects of these gauge strings
will make the description of the transition more realistic, and in particular will relate
chiral symmetry restoration to deconfinement. This restriction to the zero flux sector means that
the Hamiltonian \Eq{eq:H_AFM} describes chiral dynamics of QCD only at temperatures $T\stackrel{<}{_\sim}g^2N_c$.
As a result one should be careful when `extrapolating' the behavior of \Eq{eq:H_AFM} (or its Euclidean counterpart) to
QCD at temperatures close to the transition. This is a significant difficulty that exists in the strong
coupling approach to finite
temperature lattice QCD, which we cannot resolve here,
and that the reader should be aware of.

Nonetheless, since the model in \Eq{eq:H_AFM} shows different
critical behaviors in different large-$N$ limits, it is a sufficient
and useful choice for our study, and can tell what conditions are
needed for a large-$N$ transition to have MF critical exponents.
Also, the Hamiltonian \Eq{eq:H_AFM} is sufficient
 to understand the work of \cite{Chandrasekharan}, which also neglects excitations of the gauge-flux.

\section{The large-$N_c$ limit : a sigma model}
\label{sec:tHooft}

As mentioned in the previous section, we now fix the `t Hooft coupling $g^2N_c$
(and hence $J$) and take the large-$N_c$ limit of \Eq{eq:H_AFM}.
Writing the partition function of \Eq{eq:H_AFM} using generalized
spin coherent states gives a non-linear sigma model (NLSM) that was
first derived in \cite{RS}. The $\sigma$ field at site $\bn$ is a
$4N_f \times 4N_f$ hermitian, unitary matrix given by the $U(4N_f)$
rotation, $\sigma_\bn=U_\bn \Lambda U^{\dag}_\bn \label{sigma_n}$,
of the reference matrix $\Lambda$. In this section we work with
$m=2N_f$ that fixes
\begin{equation}
\Lambda =
\left( \begin{array}{cc} 1_{2N_f} & 0 \\ 0 & -1_{2N_f}
  \end{array}\right),
\end{equation}
and makes $\sigma_\bn$ an element of
$U(4N_f)/[U(2N_f)\times U(2N_f)]$. The action of the NLSM is
\begin{eqnarray}
A&=&\frac{N_c}2\int_0^1 dt \left[-\sum_{\bn} \Tr\Lambda U^{\dag}_{\bn}
\partial_t U_{\bn}
% \right. \nonumber \\
%&& \hskip 2cm \left.
+2x \sum_{\bn \mu} \Tr \left( \sigma_\bn \sigma_{\bn+\muhat} \right)
\right], \label{eq:action}
\end{eqnarray}
where $x\equiv J/4T$ gives the coupling in units of the temperature
$T$. Finally note that the global $U(4N_f)$ transformations are
realized here as $U_\bn \to VU_\bn$, or $\sigma_\bn \to V \sigma_\bn
V^\dag$.

For large values of $N_c$, and $x\sim O(1)$, the overall factor of
$N_c/2$ suppresses fluctuations, and the MF anstaz
\begin{equation}
\< \sigma_\bn \>= \left[ \begin{array}{cc} +\Lambda \qquad \bn\in
{\text even}, \\
-\Lambda \qquad \bn\in {\text odd}, \end{array} \right.
\label{eq:MF_NLSM}
\end{equation}
which breaks $U(4N_f)$ to $U(2N_f)\times U(2N_f)$, is exact. An
expansion around the MF ansatz, results in a chiral theory of mesons
with $O(1/N_c)$ interactions \cite{Smit}.

In contrast, if $x$ is of $O(1/N_c)$ then
 fluctuations in the second, `interaction' term, are not suppressed,
 and many sigma field configurations contribute to the path integral.
Nonetheless, the largeness of $N_c$ suppresses
 quantum fluctuations for any $x$, by making $\partial_t U \simeq 0$
 in the first, `kinetic' term.
 This can also be seen by
  rescaling the operators $Q^\eta\to N_c \, Q^\eta$ in \Eq{eq:commutation}, to find that
the commutation relations vanish at $N_c=\infty$.
  In fact, this is in analogy with the large-$S$ limit that makes
  quantum antiferromagnets into classical systems. In our case we
  get the classical Hamiltonian
\begin{equation}
{\cal H}_{\text classical}=\frac{N_c J}{4T} \sum_{\bn \mu} \Tr \left(
\sigma_\bn \sigma_{\bn+\muhat} \right), \label{eq:classical}
\end{equation}
 whose critical temperature $T_c \sim O(N_c J)$. This
   means that $x$ scales like $\frac1{N_c}\times \frac{T_c}{T}$ and therefore
   that at $T\sim O(T_c)$, $x$ is of $O(1/N_c)$. As a result the MF ansatz \Eq{eq:MF_NLSM} is not predictive at these
   temperatures, and treatments such as \cite{Umino},
   are expected to fails there. This can be also seen in
   the following way. The
   partition function of this classical system looks like
\begin{equation}
Z=\int D\sigma \, \exp \left[ -a\frac{T_c}{T} \sum_{\bn\mu}  \Tr \left(
\sigma_\bn \sigma_{\bn+\muhat} \right) \right], \label{eq:Zclassical}
\end{equation}
where $a$ is a pure number of $O(1)$ given by
$N_cJ/(4T_c)$.\footnote{For example, in
\cite{Chandrasekharan} one finds that the temperature is
 given by $T_c/N_c=1.5525(3)+ O(1/N_c)$.} In terms of $T/T_c$, \Eq{eq:Zclassical} has no $N_c$
dependence at all, which means that the
critical region has a finite width at $N_c=\infty$. Only
   outside this region will MF
   be a good description, like it is for ordinary spin systems. MF theory does not fail here since it is not
   supposed to work at all, and we believe that this is what \cite{Chandrasekharan} saw
   numerically. One might think that MF scaling fails because the ansatz used above
is temperature independent, and as emphasized above, is strictly
exact only for low $T$. In the next section we show that this
expectation is too naive.

Finally we note that the estimate of $T_c\sim N_c$ has been obtained in other strong-coupling expansions as well (for
example see \cite{DK,Chandrasekharan}). At first sight this is puzzling since there should be only one scale in this
theory which is $\Lambda_{QCD}$. This situation is reminiscent of the one discussed in \cite{Smilga}, where
the author shows that the thermal corrections to
$\< \bar{\psi} \psi \>$  in the framework of continuum chiral perturbation theory are of $O(T^2/N_c)$. Naively, this leads to
expect $T_c\sim \sqrt{N_c}$. This `paradox' is resolved in \cite{Smilga} by noting that the expansion in $T$ must have a
 finite radius of convergence of the order of the Hagedorn temperature. The latter is an external
 notion to the chiral lagrangian, and can be thought of as proliferation of string of nonzero gauge-flux.

Indeed, the action of our nonlinear sigma model \Eq{eq:action} is the chiral lagrangian in our context,
and is reached only when one neglects flux excitations (see discussion in Section~\ref{sec:Heff}).
As a consequence, it seems that the reason that in strong-coupling $T_c\sim N_c$, and not a function
of just $g^2N_c$, is the discard of flux condensation. In fact the authors in \cite{GO} study the interplay of
deconfinement and chiral symmetry restoration in strong coupling, and find that an assuming flux condensation gives a
chiral symmetry restoration temperature which is $O(N^0_c)$ at large-$N_c$. We thus expect that the inclusion flux condensation
will lead to a $T_c\sim O(N^0_c)$ in the Hamiltonian approach as well, and that it
 will be the critical behavior of the resulting system which may be indicative to what happens in the `t Hooft limit
 in the continuum.

\section{The combined large-$N_c$ and large-$N_f$ limit : Schwinger bosons}
\label{sec:Veneziano}
The aim of this section is to investigate a case where the
 MF ansatz is $T$-dependent and exact at $N=\infty$. This can be achieved by taking both $N_f$,
and $N_c$ to be large, while keeping their ratio fixed ($g^2N_c$ is
still kept fixed here). This limit was first studied by \cite{AA},
for $m=1$ with the Schwinger bosons method, and leads to a
spontaneous breakdown of $U(4N_f)\to U(1)\times U(4N_f-1)$. This
tells us that chiral symmetry may be realized differently than in
QCD, and that to break $U(4N_f)$ in the same pattern as in
Section~\ref{sec:tHooft}, one needs to generalize the procedures in
\cite{AA} to $m=2N_f$. This was done in \cite{RS1}, which found that
the {\em simplest} ansatz for general $m$ gives the same MF
equations as for $m=1$. This encourages us to choose $m=1$ as well,
and to postpone a discussion of a more sophisticated ansatz to
possible future research.

We now move to review the Schwinger bosons method, and present
 some of its results. For a more detailed discussion, we refer to
\cite{AA,Abook,Sarker} and, in the context of QCD, to \cite{thesis}.
The first step is to write
\begin{equation}
Q^\eta_\bn=\left\{\begin{array}{ll}
                        b^{\dag}_\bn \cdot M^\eta \cdot b_\bn    &       \qquad \bn \in \text{even},        \\
                        b^{\dag}_\bn \cdot \left(-M^{\eta}\right)^* \cdot b_\bn &       \qquad \bn \in \text{odd},
\end{array} \right. \label{eq:Q_eo}
\end{equation}
where the operators $b_{\alpha \bn}$ are bosonic fields that live on
the lattice sites, and have only Dirac-flavor indices. This is an
acceptable representation of $Q^\eta$, since $b_{\alpha\bn}$ obey $
\left[ b_{\alpha\bn} , b^\dag_{\beta\bmm} \right] =
\delta_{\alpha\beta} \delta_{\bn \bmm}$, and therefore
\Eq{eq:commutation} is respected. To set the representation of the
operators in \Eq{eq:Q_eo} one puts $N_c$ bosons on each site
\begin{equation}
\sum_{\alpha=1}^{4N_f} b^{\dag}_{\alpha\bn} b_{\alpha\bn} = N_c.
\label{eq:constraint}
\end{equation}
This constraint, together with the fact that the bosonic single-site
wave function is symmetric in Dirac-flavor indices, sets the
$U(4N_f)$ representation of all lattice sites to be the tableau of
Fig.~\ref{fig:Young} with $m=1$. Using \Eq{eq:Q_eo}, one can now
represent the partition function, $Z$, with bosonic coherent states,
and obtain the path integral
\begin{eqnarray}
  Z&=&\int' Db\,
Db^* \, \exp (-A) \label{eq:Z_beforeHS},\\
        A &=& -\int_0^{1/T} d\tau \,  \sum_\bn \left[ \sum_\alpha
          b^*_{\alpha\bn} \partial_{\tau}  b_{\alpha \bn}
%           \right.\nonumber
%          \\
%&&\left.
+ \frac{\bar{J}}{N_f}\sum_{\mu
           \alpha \beta}
          b^*_{\alpha \bn} b^*_{\alpha \bn+\muhat}  b_{\beta \bn}
          b_{\beta \bn+\muhat}\right],     \label{eq:S_beforeHS}
\end{eqnarray}
with $\bar{J}=J N_f/(2N_c)$. Since we keep $N_c/N_f$ and $J$ fixed,
then $\bar{J}$ is fixed as well, and we drop the bar henceforth.

The prime in \Eq{eq:Z_beforeHS} means that the path integral is
constrained to obey \Eq{eq:constraint}. Adding a Lagrange
multiplier, $\lambda_\bn$, to keep this constraint, and a
Hubbard-Stratonovich (HS) link field, $Q_{\bn\muhat}$, to decouple
the quartic interaction, one finds
\begin{eqnarray}
        A &=& \int d\tau \, \left[ \sum_{\bn \mu}
        \frac{N_f}{J}
 |Q_{\bn \mu}|^2 + i \sum_\bn  N_c \lambda_\bn
%  \right.\nonumber \\
%&-&\left.
-\sum_{\bn\alpha} b^\dag_{\alpha\bn}
 \left[\partial_{\tau} + i \lambda_\bn \right] b_{\alpha\bn} - 2{\text Re}
 \sum_{\bn \mu}  Q^*_{\bn \mu} b_{\alpha \bn} b_{\alpha
   \bn+\muhat}\right]. \nonumber  \\   \label{eq:S_HS}
\end{eqnarray}
An integration over the bosons, gives the effective
action for $Q$ and $\lambda$,
\begin{eqnarray}
A_{\text eff}&=&4N_f\left\{\int d\tau \, \left[ \sum_{\bn \mu} \frac{1}{4J}
|Q_{\bn
    \mu}|^2 + i \sum_\bn  \kappa \lambda_\bn\right]
%     \right. \nonumber
%\\
%&&\left. \hspace{1cm}
 + \tr \log G^{-1} \right\}, \label{eq:Seff}
\end{eqnarray}
where $\kappa=N_c/(4N_f)$, and where $G$ is the propagator of a {\em
single} boson, such that $\log \det G \sim O(1)$. For large values
of $N_f$, but $\kappa\sim O(1)$, and $T/J\sim O(1)$, fluctuations
are suppressed around any stable MF ansatz. The action of the ansatz
$Q_{\bn \mu}(\tau)=Q, \lambda_\bn= i \lambda$, is
\begin{eqnarray}
A_{\text{MF}} &=& \frac{4N_fN_s}{T} \left[ \frac{dQ^2}{4J} -\lambda
\left(\kappa + \frac{1}{2}\right)
%\right. \nonumber \\
%&& \left.
+ \frac{2}{N_s\beta} \sum_\bk \log
2\sinh \left( \frac{\beta \omega_\bk}{2} \right)\right],
\label{eq:S_MF1}
\end{eqnarray}
where $\beta=1/T$. Also, $N_s$ is the number of sites, $\bk$ takes
values in the Brillouin zone of the even sublattice. Defining
$\gamma=\frac1{d}\sum_{i=1}^d \cos (k_i/2)$, one finds that the
function $\omega_\bk=\sqrt{\lambda^2-4d^2Q^2\gamma^2_\bk}$ becomes
$\omega^2_\bk\simeq \Delta^2+c^2\bk^2$ at low $\bk$, with the mass
gap $\Delta=\sqrt{\lambda^2-4d^2Q^2}$, and with $c^2=dQ^2$. A
minimization of \Eq{eq:S_MF1} with respect to $\lambda$ and $Q$,
gives the MF equations
\begin{eqnarray}
\frac{4Q}{dJ}&=&\frac2{N_s} \sum_{\bk} \frac{Q\gamma^2_\bk}{\omega_\bk} \left( n_B(\omega_\bk) + \frac12 \right), \label{eq:MF1} \\
\kappa+\frac12 &=& \frac2{N_s} \sum_\bk \frac{\lambda}{\omega_\bk} \left( n_B(\omega_\bk)+\frac12 \right) . \label{eq:MF2}
\end{eqnarray}
with $n_B(\omega)=(e^{\beta\omega}-1)^{-1}$. These equations are exact
at large-$N$, and their solution gives the large-$N$ phase
diagram.
This is different from Section~\ref{sec:tHooft} where the
mean-field ansatz is exact only at low $T$. In fact, we believe
that the Schwinger bosons approach is an appropriate candidate to be the
`finite $T$ mean field theory' mentioned in
\cite{Chandrasekharan}. Another candidate is presented in
Section~\ref{sec:largeNf}.

The point at which the inverse correlation length $\Delta$ vanishes,
signals the condensation of the bosons, which breaks $U(4N_f)$ to
$U(1)\times U(4N_f-1)$. Investigating
Eqs.~(\ref{eq:MF1})--(\ref{eq:MF2}) shows that this happens for
$T=0$ at $\kappa \ge \kappa_c$,\footnote{$\kappa_c$ is $\sim 0.19$
for $d=2$ \cite{AA}, and $\sim 0.0778$ for $d=3$ \cite{Abook}.} and
that the symmetry is restored at $T=T_c(\kappa)$, which is finite
\cite{AA,Sarker}.

To extract the critical exponents, one expand
Eqs.~(\ref{eq:MF1})--(\ref{eq:MF2}) around $T_c$. We illustrate this
in Appendix~\ref{app:schwingerscaling}, where we redo the
calculation of \cite{Sarker} to show that MF scaling holds only for
$d>4$, and that in terms of the reduced temperature $t=|T-T_c|/T_c$,
one finds that
$\Delta\sim t^{1/2}$. For $2<d\le 4$, infrared (IR)
fluctuations change this behavior to $t\sim -\Delta^2\log \Delta$
for $d=4$, and to $\Delta\sim t^{1/(d-2)}$ if $2<d<4$, which
coincides with the behavior of the $CP_N$ model at large-$N$
\cite{CPN}. Extending these results, we also show that the critical
region for $2<d\le4$ is nonzero and depends on $N_f$ and $N_c$ only
through $\kappa$.

We conclude this section by noting that the overall factor of $4N_f$
in \Eq{eq:Seff} does not contradict the fact that IR modes are not
suppressed. The simple reason is that the fields $Q$ and $\lambda$
are $U(4N_f)$ singlets, and are not the order parameters of the
transition. This means that \Eq{eq:Seff} {\em is not} the LGW action
and that fluctuations in the order parameters need not be
suppressed. Large-$N_f$ suppresses fluctuations in $Q$ and
$\lambda$, which means that the results one obtained here are exact
at $N_f=\infty$. Taking these fluctuations into account leads to
corrections of $O(1/N_f)$ to the critical exponents, and to $T_c$.

\section{The large-$\boldmath{N_f}$ limit: fermions}
\label{sec:largeNf}

Here we take $N_c=1$ and work in two spatial dimensions. We begin by
using \Eq{eq:M_eta} to write \Eq{eq:H_AFM} as
\begin{equation}
H=-\frac{\bar{J}}{N_f}\sum_{\bn\mu\atop
\alpha\beta}\psi^{\dag\alpha}_{\bn}
\psi^{\alpha}_{\bn+\muhat}\psi^{\dag\beta}_{\bn+\muhat}\psi^{\beta}_{\bn}+\frac{\bar{J}}{N_f}\sum_{\bn\alpha}
\psi^{\dag\alpha}_\bn \psi^{\alpha}_\bn, \label{eq:H_fermi}
\end{equation}
where we define $\bar{J}=JN_f/2\sim N_f/g^2$. Since we take
large-$N_f$ with fixed $g^2/N_f$ then
 $\bar{J}$ is fixed as well, and for brevity, we drop the bar
henceforth. To set
the representation of the operators in \Eq{eq:Q} we follow
\cite{AA,RS1} and put $2N_f$ fermions on each site
\begin{equation}
\sum_{\alpha} \psi^{\dag\alpha}_\bn \psi^\alpha_\bn = 2N_f.
\label{eq:constraint1}
\end{equation}
This also means that the second term in \Eq{eq:H_fermi} is a
constant which we drop. Since the wave function of the fermions is
{\em anti}-symmetric in the Dirac-flavor indices, then
\Eq{eq:constraint1} puts it in the representation given in
Fig.~\ref{fig:Young} with $m=2N_f$. The partition function of
\Eq{eq:H_fermi} is \cite{RS}
\begin{eqnarray}
Z&=&\int_{-\pi T}^{\pi T} D\lambda \int DQ DQ^* D\psi
D\bar{\psi} \, e^{-A}, \\
A&=&\int d\tau \left\{
 \sum_\bn \bar{\psi}_\bn \left( \partial_\tau - i \lambda_\bn
\right) \psi_\bn
% \right. \nonumber \\
%&&\left. \hspace{1cm} +
\sum_{\bn,\mu} \left(\bar{\psi}_\bn Q_{\bn\muhat}
\psi_{\bn+\muhat}+ h.c. \right)
% \right. \nonumber \\
%&&\hskip 0.5cm   \left.
+ \frac{N_f}{J} \sum_{\bn,\mu} |Q_{\bn\muhat}|^2 + i\, 2N_f \sum_\bn
\lambda_\bn \right\}. \nonumber \\  \label{eq:Afermi}
\end{eqnarray}
As in the previous section, $\lambda_\bn$ is a constraint field that
keeps \Eq{eq:constraint1} on each site, and $Q_{\bn\muhat}(\tau)$ is
a HS field that decouples the four-Fermi interaction of
\Eq{eq:H_AFM}. Apart from the global $U(4N_f)$ symmetry,
\Eq{eq:Afermi} is also invariant under the $U(1)$ gauge
transformation $\psi_\bn\to e^{i\Lambda_\bn(\tau)} \psi_\bn,
\bar{\psi}_\bn\to \bar{\psi}_\bn e^{-i\Lambda_\bn(\tau)},
Q_{\bn\muhat}\to
e^{i(\Lambda_\bn(\tau)-\Lambda_{\bn+\muhat}(\tau))}Q_{\bn\muhat},\lambda_\bn\to
\lambda_\bn+\partial_\tau \Lambda_\bn$. Demanding that the fields
$Q,\psi,\bar{\psi}$ will be single-valued, and that $\lambda_\bn\in
[-\pi T, \pi T]$ remains time-independent, restricts
$\Lambda_\bn(\tau)$ to be time-independent as well.

\begin{figure}[htb]
\includegraphics[width=6cm]{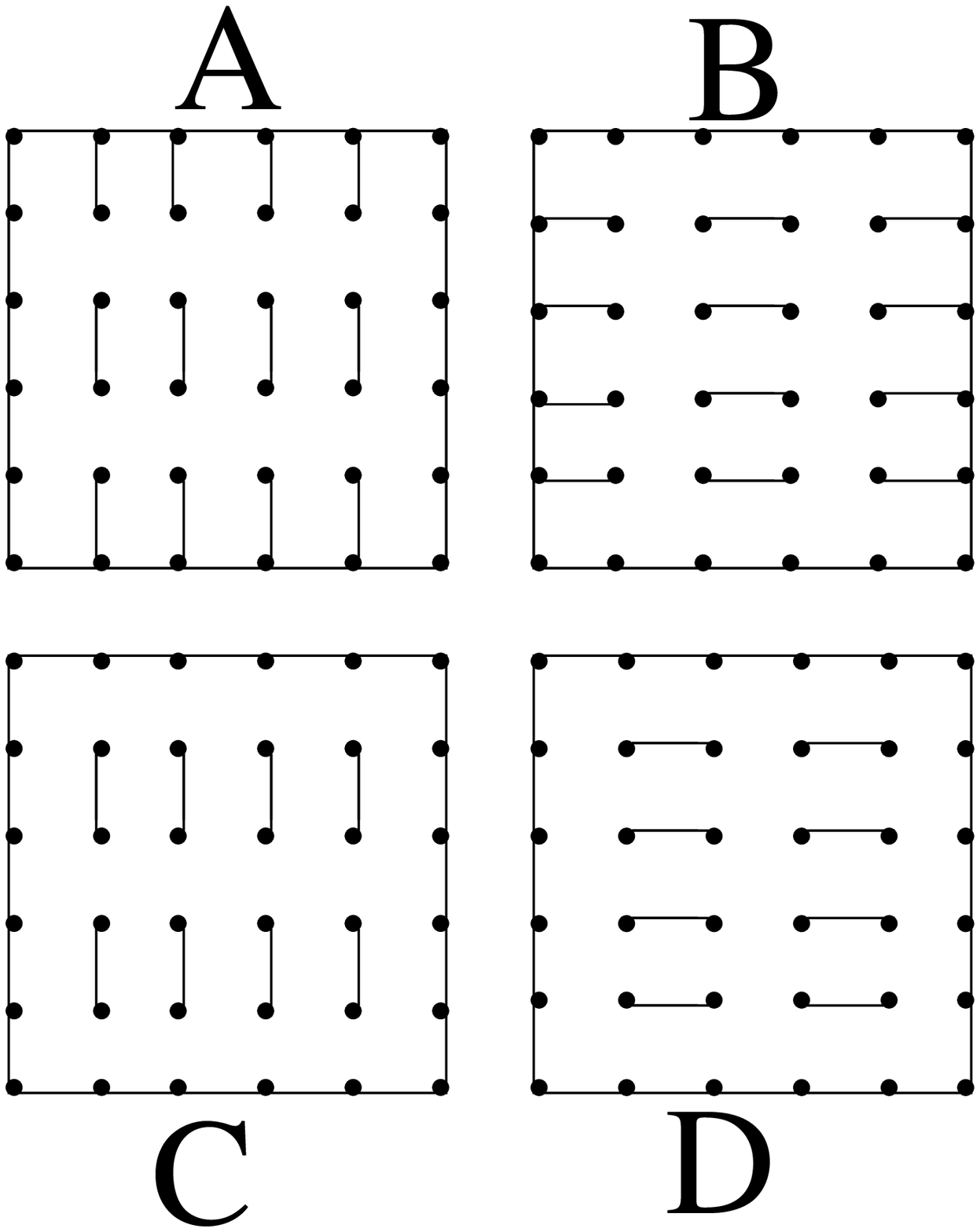}
\caption{The four degenerate
  spin-Peierls ground states, which break translation and rotation symmetries. The low
  energy effective action we compute begins from state ``A''.
\label{fig:spinPeierls}}
\end{figure}

To proceed, one integrates over the fermions and obtains the action
\begin{eqnarray}
A_{\text eff}&=&4N_f\left\{\int d\tau \left[ \frac{1}{4J} \sum_{\bn,\mu}
|Q_{\bn\muhat}|^2 +  \frac{i}2 \sum_\bn \lambda_\bn \right]
%\right. \nonumber \\
%&&\left. \hspace{1cm}
+ \tr \log
D^{-1}(Q,\lambda)\right\},
\label{eq:Afermi1}
\end{eqnarray}
where $D(Q,\lambda)$ is the Dirac operator of a {\em single}
fermion, such that $\tr \log D \sim O(1)$, and can be read from
\Eq{eq:Afermi}. For $T/J\sim O(1)$, the overall factor of $4N_f$ in
\Eq{eq:Afermi1} suppresses fluctuations in $Q$ and $\lambda$, and
one can exactly solve for the ground state which was shown to be the
spin-Peierls state \cite{MA1,MA2,RS}. It has $\lambda_\bn=0$ on all
sites, and $|Q_{\bn\muhat}(\tau)|=q$ on a subset of the lattice
links ${\cal B}$, but zero otherwise. This ground state is four-fold
degenerate, and is shown pictorially in Fig.~\ref{fig:spinPeierls}.
It is clear that this state breaks lattice translations and
rotations, which are then restored at the critical temperature
$T_c$. To discuss this transition, one writes the MF effective action
for the spin-Peierls state
\begin{equation}
A_{MF}=\frac{4N_f N_s}{T}\left[ \frac{1}{4J}q^2 \times \frac14
\times 2 +\frac1{2\beta} \log n_F(q) n_F(-q)\right], \label{eq:A_MF}
\end{equation}
where $\beta=1/T$, and $n_F(\epsilon)=(e^{\beta\epsilon}+1)^{-1}$. A minimization of
\Eq{eq:A_MF} with respect to $q$ yields the MF equation
\begin{equation}
q/2J=\tanh (q/2T),
\end{equation}
that describes a stable ordered phase with $q>0$ below $T_c=J$, and
a MF scaling $q\sim (T_c-T)^{1/2}$ in its vicinity. Here we reach
the same `puzzle' as did
\cite{Rosenstein,Kogut2,Pelissetto1,Pelissetto2} for the
Gross-Neveu, Yukawa and classical 2D Heisenberg models: the result
above is {\em exact} at $N_f=\infty$, but for any finite $N_f$ it is
wrong. There, the scaling of $q$ is dictated by the symmetry
breakdown, and by $d$. The resolution of this puzzle is that the
critical region shrinks with $N_f$
\cite{Kogut2,Pelissetto1,Pelissetto2}. We now turn to show how this
happens for our model.

We begin by calculating the effective action for fluctuations by
replacing $\lambda_\bn\to \frac1{\sqrt{4N_f}}\lambda_\bn$, and
$Q_{\bn\muhat}\to q+\frac1{\sqrt{4N_f}}Q_{\bn\muhat}$ in $A_{\text
  eff}$ (Here $q$ is the solution of the MF equation, and is nonzero only on
${\cal B}$). We expand \Eq{eq:Afermi1} to $O(1/N_f)$ in
Appendix~\ref{app:bubbles}, where we work in Matzubara space and
show that the masses $m$ of the zero Matzubara fields $\phi\equiv
Q(\omega=0)$ vanish at $T_c$ like $m^2\sim t \equiv |T_c-T|/T$. All
other fields are massive, and we proceed by integrating them out in
Appendix~\ref{app:Intmassive}. The effective action we find is
\begin{eqnarray}
A^{0}_{\text eff}&=&\sum_{\bn \mu} \frac12 m^2_{\bn \mu} |\phi_{\bn
}|^2 + \sum_{\bn_1 \bn_2 \mu} v_{\bn_1 \bn_2 \mu} \, {\text Re} \,
\phi_{\bn_1 \muhat} \phi_{\bn_2 \muhat} \nonumber  \\
&+& \frac1{\sqrt{N_f}} \sum_{
\{ \bn \}, \mu\nu} \left[ V^{(3)}_{\mu\nu}(\{ \bn \})
 \phi^*_{\bn_1 \nu} \phi^*_{\bn_2 \mu} \phi_{\bn_3 \mu}  +
 h.c. \right] \nonumber
\\
&+&\frac1{N_f} \sum_{\{\bn\}, \mu\nu} \left[
V^{(4)}_{\mu\nu}(\{\bn\}) \phi^*_{\bn_1 \mu} \phi^*_{\bn_2 \nu}
\phi_{\bn_3 \mu} \phi_{\bn_4 \nu} + h.c. \right], \nonumber \\ \label{eq:A_eff_zero_T-}
\end{eqnarray}
with $m^2,v\sim O(t)$, $V^{(3)} \sim O(t^{1/2})$, and $V^{(4)}\sim
O(1)$ (detailed expressions for these are given in
Appendix~\ref{app:Intmassive}.)

As emphasized by \cite{Kogut2,Pelissetto1,Pelissetto2} it is now
clear where does the puzzle come from. Taking $N_f\to \infty$ before
$t\to 0$ gives a Gaussian model. Switching the order of the limits,
we get a weakly coupled $\phi^3$ theory, and the
four-fold degeneracy of the spin-Peierls state leads one to expect
that the universality class of the transition is of a $Z_4$ model in
$d=2$ (see for example \cite{Senthil} and its references). This
points to a crossover behavior, where the susceptibility $\chi$
diverges like
\begin{equation}
\chi\sim t^{-1} f(x), \hspace{1cm} x=t N^p_f, \label{eq:f}
\end{equation}
and where $f(x)$ is a scaling function that determines the critical
behavior.

\begin{figure}[htb]
\includegraphics[width=3cm]{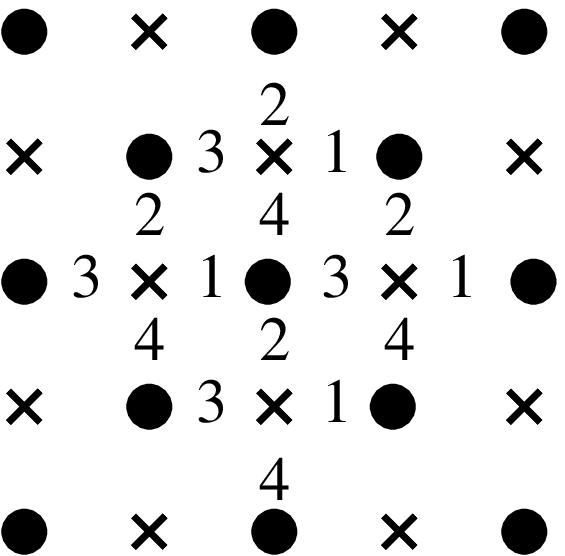}
\caption{The lattice of even sites, denoted by X, and odd sites,
denoted by $\bullet$. Each even site has four link fields denoted
by $\phi_i$, with $i=1,2,3,4$. \label{fig:fcc_lattice}}
\end{figure}

A calculation of $f(x)$ is simpler in the high temperature phase,
where $q=0$. There, the effective action has only diagonal and
degenerate mass terms, and no cubic interactions. In momentum space
the action reads
\begin{eqnarray}
A^{0}_{\text eff}&=&\sum_{\bk i} \frac12 m^2 |\phi_{\bk i}|^2
%\nonumber \\
%&+&
+\frac1{N_f} \sum_{\bk_1 \bk_2 \atop \bk_3 \bk_4} \sum_{i j
\atop k l}
 \left[ V_{ijkl}(\{\bk\}) \phi^*_{\bk_1 i} \phi^*_{\bk_2 j} \phi_{\bk_3 j}
 \phi_{\bk_4 l}  +  h.c. \right], \nonumber \\ \label{eq:A_eff_zero_T+}
\end{eqnarray}
where the momentum $\bk$ belongs to the first Brillouin zone
 of the even sublattice, and the index $i=1,2,3,4$ denotes the four links
 outgoing from each even site, see Fig.~\ref{fig:fcc_lattice}.
In Appendix~\ref{app:Intmassive} we show that $m^2=(T-T_c)/(4TT_c)$,
and give the form of $V\sim O(1)$.

We calculate the susceptibility
$\chi_{ij}(\bk,\bk')=\<\phi_i(\bk)\phi^*_j(\bk')\>$
 to leading order in $1/N_f$ in
Appendix~\ref{app:f_of_x}, and show that it coincides with \Eq{eq:f}
if $x=t\, \sqrt{\frac{6}{13}}\sqrt{N_f}$, and $f(x)=1-1/x^2$. This
tells us that only if $x=\infty$ then one obtains the MF scaling of
$\chi\sim t^{-1}$. As $x$ decreases from infinity, this behavior
changes, and will eventually be determined by the nearest fixed
point of the $\phi^4$ theory of \Eq{eq:A_eff_zero_T+}.\footnote{Or
to a possible modification of that by topological effects, induced
by the compactness of the gauge field. These have been shown to be
important in the $T=0$ phase transition between the spin-Peierls and
N\'eel states \cite{Senthil}.} This
crossover behavior will occur when $x\sim O(1)$ or $t\sim
O(1/\sqrt{N_f})$.

To make a connection with the results of \cite{Kogut2}, we recall
that a $\phi^4$ theory in $d$ dimensions and a $O(1/N)$ coupling
enters its critical region when its mass squared is
$m^2\stackrel{<}{_\sim} (1/N)^{\frac2{4-d}}$. Our result fits here
if we put $d=0$, and gives the largest critical region possible. The
reason is that here the bare propagator of $|\phi|$ is momentum
independent, and is simply given by $(\frac12 m^2)^{-1}$. This
degeneracy will be removed by higher orders in $1/N_f$, and can be
avoided by analyzing other ground states, such as the flux phase
\cite{MA2}. This also means that in a generalization to other values
of $N_c$ and $d$, it is reasonable to expect that the width of the
critical region will be given by $t\sim 1/N^p$, with $p\ge 1/2$, or by
a possible logarithmic modification of that.

In contrast to the combined limit of large-$N_c$ and large-$N_f$, the critical region is
suppressed here because the order parameter of the spin-Peierls state is the $Q$ field, whose action has an overall factor of
$4N_f$. The latter suppresses the interaction terms in the LGW
action given in \Eq{eq:A_eff_zero_T+}, and can make MF scaling
exact.\

\section{Summary}
\label{sec:summary}

We work with an effective Hamiltonian derived from the lattice QCD Hamiltonian in
the strong coupling expansion at second order.  The large-$N_c$ limit
of \Eq{eq:H_AFM} is the analog of the large-$S$ limit of quantum
antiferromagnets. For the latter, quantum fluctuations are
suppressed, and one is left with a classical magnet, which is by no
means a simple object. In particular, its
 critical behavior cannot be approached with a MF spin ansatz, because
 the latter is a good approximation only at sufficiently low
 temperatures. The spin $S$
  determines the energy scale of the system, and for our case of
  \Eq{eq:H_AFM}, this leads to $T_c\sim N_c$, and to the fact that in
  terms of $T/T_c$, the critical
region of our model is finite at $N_c=\infty$. We suggest that this
   is what \cite{Chandrasekharan} saw numerically in the action
   formalism, and that MF does not fail there, since it was not
   supposed to work in the first place.

   The combined limit of large $N_c$ {\em and} large-$N_f$, but fixed
   ratio $N_c/N_f$, is more convenient to study the
   transition. There, the critical temperature is finite, and the
  MF ansatz, which is exact at $N_f=N_c=\infty$, is temperature
 dependent, and describes the
   critical region exactly. For $d>4$, it gives Gaussian
   scaling, but for $d\le 4$, IR modes change this and one can
   calculate the exponents exactly. Also, the width of the critical region depends
   on $N_f$ and $N_c$ only through $N_c/N_f$, and is
   nonzero.

\begin{figure}[htb]
\includegraphics[width=6cm]{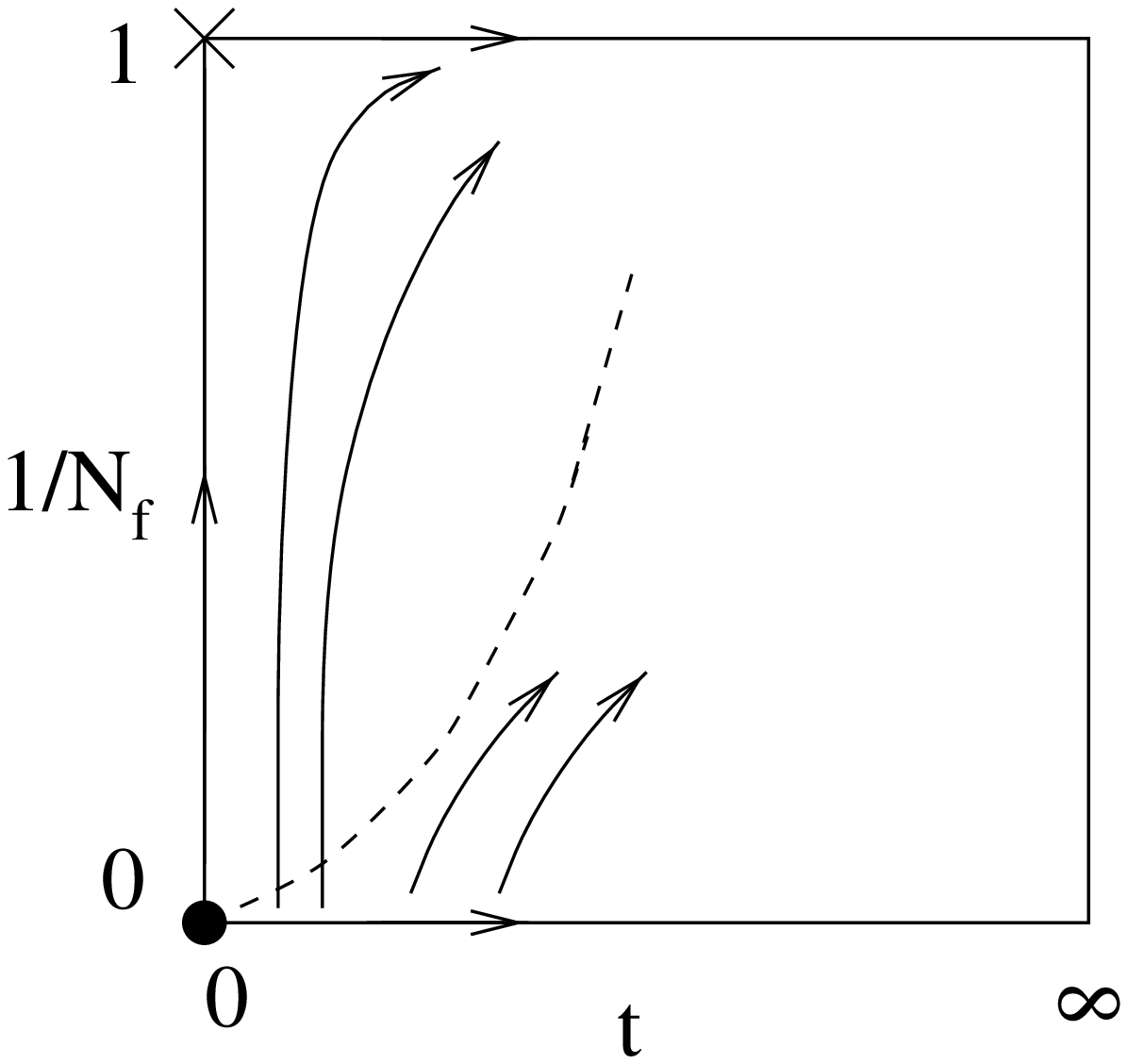}
\caption{A cartoon of a possible renormalization group (RG) flow in
  the $(t,1/N_f)$ space ($t$ is the reduced temperature
  $\frac{|T_c-T|}{T_c}\, $.) RG
drives the system away from the Gaussian fixed point denoted by the
`$\bullet$'. This happens along the $1/N_f$ axis towards a possible
nontrivial fixed point denoted by the `X' (which for simplicity we
placed at $N_f=1$). The critical region is to the left of the dashed
line, $1/N_f \sim t^2$. \label{fig:crossover}}
\end{figure}

   In the large-$N_f$ limit, we have studied the case of $d=2$, and $N_c=1$.
   Here the MF ansatz is also exact for all $T$,
and breaks lattice symmetries.
 This transition is characterized
   by MF critical exponents if $N_f$ is sent to infinity {\em before}
   sending $t=|T-T_c|/T_c$ to zero. In contrast, one can take
   $N_f\to \infty$, and $t\to 0$ with fixed $x\sim t\sqrt{N_f}$. This leads to a
   crossover from a Gaussian fixed point (at $x=\infty$) to
   a nontrivial fixed point of a weakly coupled scalar field theory.
   A schematic view of a possible renormalization group
   flow for this case is given in
   Fig.~\ref{fig:crossover}.

To conclude, it is clear that the critical region in large-$N$ phase
transitions is suppressed when the Landau-Ginzburg-Wilson action
 has an overall factor of $N^\alpha$ with $\alpha>0$.\footnote{$\alpha=1$ in conventional cases where
 $N$ is related to a global symmetry, while for deconfinement in a pure
 $SU(N)$ gauge theory one expects $\alpha=2$.} Taking the
 $N\to \infty$ can then remove IR fluctuations
  in the order parameter, and makes Gaussian scaling exact. We stress that this does not happen for all
large-$N$ theories, and for our model Hamiltonian of \Eq{eq:H_AFM},
it happens only in the limit of large-$N_f$ and fixed $N_c$.

\section{Future prospects}
\label{sec:future}

A generalization of
     the large-$N_f$ discussion to $d=3$, and $N_c>1$ is
    straight forward, but relatively costly. For example, one has
     to find the true vacuum out of all possible ansatze (although an effective action for fluctuation
    around a metastable vacuum can also be calculated). We
    believe that the results of such a generalization will differ from the case
    studied in Section~\ref{sec:largeNf} only in detail, and not in
    principle.\footnote{For example, one expects that both $x$ and
      $f(x)$, depend on $t$ and $N_f$, in a way that changes with the
      dimension $d$.}
 In view of
    that, and the limited relevance of the large-$N_f$ limit for QCD in our context,
    we find no point to do so analytically. A numerical check of our
    predictions, and of the possible generalization mentioned above, can be made
    with similar methods to the one of
    \cite{Chandrasekharan}. Ideally this should be complemented with an analytical
    study in the action formalism at large-$N_f$, in similar lines to
    Section~\ref{sec:largeNf}.

    As suggested by \cite{Smilga} and \cite{Kogut2} the critical region may be suppressed in planar QCD
    (large-$N_c$ with fixed
    $g^2N_c$ and fixed $N_f$). While \cite{Kogut2} suggests that the outside the critical region,
     one observes a MF behavior, then
  \cite{Smilga} expects that the chiral condensate will not change as a function of $T$ at
  all until the critical region is reached.
   It will be
    interesting to check these theoretical expectations, but unfortunately the current study of our model Hamiltonian
     cannot (and is not intended to) accomplish this. The reason
    is two-fold; First we are
    very far from the continuum limit. Second, the Hamiltonian \Eq{eq:H_AFM} is not designed to explore
    temperatures which
    are close to the deconfinement temperature. While the latter problem can be in principle tackled by
    including gauge-flux condensation (see the discussion at the end of
    Section~\ref{sec:tHooft}), it is hard to see how to overcome the former.
    Approaching continuum physics with numerical simulations at large-$N_c$ and/or large-$N_f$ is very costly.
    While simulating dynamical quarks at large-$N_c$ is unrealistic, then quenching them or using dynamical fermions
    at large-$N_f$ is expensive.

    This leads us to seek phase
    transitions in other related
    models in which the critical region may be suppressed as well. An obvious choice, which
    is numerically `cheap',
     is deconfinement phase transitions
    in large-$N_c$, where the role of the LGW
    action is played by a Polyakov loop action, in the spirit of
    \cite{P_loops}.
    The fact that the pure gauge theory
has a first order deconfinement transition at large-$N_c$ (e.g. see
\cite{MikeFeb05}) means that to study such a phenomenon in pure
gauge, one has to approach a hidden Hagedorn transition, where the
mass of the loop vanishes. This was done in \cite{Hagedorn_paper},
but the results are not accurate enough to unambiguously determine
the critical exponents.\footnote{The reason is the high
calculational cost, and the fact that the field configurations used
there are meta-stable rather than stable.}

Adding bosonic matter to the pure gauge theory may change the
deconfinement transition
into second order. In particular, using bosons in a bi-fundamental
representation with $N_c$ even, leaves a $Z_2$ subgroup of the $Z_N$
center intact. As a result, the deconfinement transition (if second order) is in
the Ising model universality class, and by adding a flavor structure to the
bosons, one can study the chiral symmetry restoration as well \cite{Sannino}. Also, these scalar
theories are interesting on their own right, as they may serve to
check the large-$N_c$ relation between supersymmetric and
non-supersymmetric gauge theories discussed in \cite{ASV}.

\appendix

\section{The critical exponents and the width of the critical region
  in the combined large $\boldmath{N_c}$ and large $\boldmath{N_f}$ limit}
\label{app:schwingerscaling}

To extract the critical exponents from
Eqs.~(\ref{eq:MF1})--(\ref{eq:MF2}),  we restrict to the
disordered phase, where there are no Bose condensates, and where one can
replace the momentum sums in Eqs.~(\ref{eq:MF1})--(\ref{eq:MF2})
with the integrals
\begin{eqnarray}
\Lambda&=&\int \left(\frac{dk}{2\pi} \right)^d \frac{\gamma^2\,
\coth(y\,\sqrt{
1-\gamma^2+\gamma^2\Delta^2})}{\sqrt{1-\gamma^2+\gamma^2\Delta^2}},
\nonumber \\ \label{eq:MF1int} \\
2\kappa+1 &=& \int \left(\frac{dk}{2\pi} \right)^d
\frac{\coth(y\,\sqrt{
1-\gamma^2+\gamma^2\Delta^2})}{\sqrt{1-\gamma^2+\gamma^2\Delta^2}} .
\label{eq:MF2int}
\end{eqnarray}
Here we also divided \Eq{eq:MF1} by $Q/\lambda$, and defined
$y=\beta\lambda/2$, and $\Lambda=8\lambda/Jd$. Let us now concentrate
on \Eq{eq:MF2int}, and on values of $\kappa$ where the $T=0$ ground
state is ordered. For each such $\kappa$ there exists
$y=y_c(\kappa)$, where
 \Eq{eq:MF2int} gives $\Delta=0$, signaling the transition.

Following \cite{Sarker} we denote the integral on the right hand side
of \Eq{eq:MF2int} by $I(y,\Delta)$, and write \Eq{eq:MF2int} as
\begin{eqnarray}
I(y,0)-I(y_c,0)&=& \int \left(\frac{dk}{2\pi} \right)^d \left[
\frac{\coth(y\,\sqrt{
      1-\gamma^2})}{\sqrt{1-\gamma^2}}- \right. \nonumber \\
&&\left. \frac{\coth(y\,\sqrt{
      1-\gamma^2+\gamma^2\Delta^2})}{\sqrt{1-\gamma^2+\gamma^2\Delta^2}}\right].
       \label{eq:scaling1}
\end{eqnarray}
For $I'\equiv - \partial I(y,0) /\partial y$ we find
\begin{equation}
I'=\int \left(\frac{dk}{2\pi}\right)^d \frac{1}{\sinh^2y\sqrt{1-\gamma^2}}>0,
\end{equation}
which means that for $y\simeq y_c$, the left hand
side scales like $(y_c-y)>0$.
Next, we expand the right hand side of \Eq{eq:scaling1} in $\Delta$,
and note that sufficiently close to $T_c$ the integral is controlled
by IR momenta. For these we write $(1-\gamma^2)=\bk^2/4d$ to find
that the IR part of the integral is
\begin{equation}
f(\Delta)=\frac{(4d\Delta)^2}{y}\int_0^R
\left(\frac{dk}{2\pi} \right)^d \frac{1}{\bk^2(\bk^2+4d\Delta^2)}.
\label{eq:f_delta}
\end{equation}
Here $R$ separates between the IR modes and rest, and we take $1\gg
R\gg 2\Delta\surd d$. As a result \Eq{eq:scaling1} becomes
\begin{equation}
(y_c-y)/y_c = A_d \left[ \begin{array}{lc} \Delta^2 & \hspace{1cm} d>4,\\
    \Delta^2 \log \frac{R}{4 \Delta} &   \hspace{1cm} d=4,\\
    \Delta^{d-2} & \hspace{1cm} 2<d<4, \end{array} \right. \label{eq:scaling3}
\end{equation}
with the finite $A_d$ given by
\begin{equation}
A_d = \frac{(4d)^2I'}{y_c^2}\left[ \begin{array}{lc} {\displaystyle \int
    \left(\frac{dk}{2\pi} \right)^d \frac{1}{\bk^4}}  &  \hspace{0.2cm} d>4,\\
    {\displaystyle \frac1{(2\pi)^4}} &  \hspace{0.2cm} d=4,\\
    {\displaystyle \int^{R/\Delta}_0 \left(\frac{dx}{2\pi} \right)^d \frac{1}{x^2(x^2+4d)}} &  \hspace{0.2cm} 2<d<4, \end{array} \right. \label{eq:Ad}
\end{equation}

We proceed to write \Eq{eq:MF1int} as
\begin{eqnarray}
\Lambda(1-\Delta^2)&=&2\kappa+1 - \int \left(\frac{dk}{2\pi} \right)^d
\sqrt{1-\gamma^2+\Delta^2\gamma^2} \times \nonumber \\
&&\coth(y\,\sqrt{1-\gamma^2+\Delta^2\gamma^2}), \label{eq:g_delta}
\end{eqnarray}
which at $T=T_c(\kappa)$ gives
\begin{equation} \Lambda_c=2\kappa+1 - \int
\left(\frac{dk}{2\pi} \right)^d \sqrt{1-\gamma^2}
\coth(y_c\,\sqrt{1-\gamma^2}). \label{eq:g_delta0}
\end{equation}
This means that
\begin{equation}
 \Lambda_c-\Lambda=-\frac12\Delta^2(y_cI'''+\Lambda_c)+(y_c-y)I'',
\end{equation}
with
\begin{eqnarray}
I''&=&\int \left(\frac{dk}{2\pi}\right)^d \frac{1-\gamma^2}{\sinh^2 (y_c \sqrt{1-\gamma^2})},\\
I'''&=&I'-I''.
\end{eqnarray}
Together with $T=\Lambda Jd /(16y)$, this gives
\begin{equation}
t\equiv (T-T_c)/T_c=B_d\Delta^2 + A'_d \times \left[ \begin{array}{lc} \Delta^2 & d>4,\\
    {\displaystyle \Delta^2 \log \frac{R}{4 \Delta}} & d=4,\\
    \Delta^{d-2} & 2<d<4, \end{array} \right. \label{eq:scaling4}
\end{equation}
with
\begin{eqnarray}
B_d&=&\frac12 (1+\frac{Jd}{16T_c} \, I'''), \\
A_d'&=&A_d(1-\frac{Jd}{16T_c}I'').
\end{eqnarray}
We choose to work with values of $\kappa$, where both $A_d'$, and
$B_d$ are positive.\footnote{Numerically, we find that for
$\kappa\stackrel{>}{_\sim}0.21$ then $A'_3$ is negative for $d=3$. This leaves
us with the window $0.0778<\kappa<0.21$ to explore the symmetry
breakdown. We are not
aware of any discussions in the literature with regards to the cases
where $A'_d<0$.} Also note the $B_d$ should include other
  $O(1)$ coefficients that come from momenta
  $\bk^2>R^2$ in the left hand side of \Eq{eq:scaling1}.

For sufficiently small $\Delta$, \Eq{eq:scaling4} leads to $\Delta\sim
t^{1/2}$ for $d>4$, while for $d=4$ and $2<d<4$, one finds that
$-t\sim \Delta^2\log \Delta$, and $\Delta\sim t^{1/(d-2)}$,
respectively. The width of the critical region is given by comparing
the two terms in the right hand side of \Eq{eq:scaling4}, and we
find that MF behavior will
 fail when
\begin{equation}
\Delta/R\ll \frac1{4} e^{-B_4/A'_4}
\end{equation}
for $d=4$, and when
\begin{equation}
\Delta\ll \left(\frac1{B_d/A'_d}\right)^{1/(4-d)}
\end{equation}
for $2<d<4$. Since $B_d/A'_d$ depends only on $\kappa$ we conclude
that the critical region of the phase transition in the large $N_c$
and large $N_f$ limit does not shrink with $N_f$ or $N_c$, and is
nonzero (Here we ignore scenarios where one tunes $\kappa$ to
 special values where $B_d/A'_d\gg 1$, and concentrate in generic
 values of $\kappa$ where $B_d/A'_d\sim O(1)$.)

\section{Calculation of $A_{\text eff}$ in the large-$N_f$ limit}
\label{app:bubbles}

In this section we calculate the large-$N_f$ effective action for
the auxiliary fields $Q$ and $\lambda$ around $T_c$, where the
spin-Peierls state dissolves.
 We begin by replacing $Q_{\bn\muhat}\to
q+\frac1{\sqrt{4N_f}}Q_{\bn\muhat}$, and $\lambda_\bn\to
\frac{1}{\sqrt{4N_f}}\lambda_\bn$ in \Eq{eq:Afermi1}, and move to
Matzubara space, where we denote the frequencies of the fermions by
$\epsilon$, and the external bosonic frequencies by $\omega$. The
action we get is
\begin{eqnarray}
A&=&A_{MF}+\sqrt{4N_f}\left[ \frac{\sqrt{\beta}}{2J}\sum_{\bn \mu \in
{\cal B}} q \, {\text
    Re} \, Q_{\bn \muhat}(\omega=0) +
%     \right. \nonumber \\
%&&\left.
\beta\frac{i}2 \sum_\bn \lambda_\bn
\right] + \frac{1}{4J} \sum_{\bn \mu}\sum_\omega
  |Q_{\bn \muhat}(\omega)|^2 + \delta A, \nonumber \\ \hspace{-2 cm}
  \label{eq:A_eff0}
\end{eqnarray}
\begin{eqnarray}
\exp(-\delta A)&=& \< \exp \left\{ \frac{i}{\sqrt{4N_f}} \sum_\bn \lambda_\bn
  \sum_\epsilon \bar{\psi}_\bn(\epsilon) \psi_\bn(\epsilon)
%   \right. \nonumber \\
%&& \left.
-\frac1{\sqrt{4N_f\beta}}
  \sum_{\bn \mu \atop \epsilon \omega} \left[
    \bar{\psi}_\bn(\epsilon+\omega) Q_{\bn\muhat}(\omega)
    \psi_{\bn+\muhat}(\epsilon)
%    \nonumber \right. \right. \\
%&&\left. \left.
+ h.c. \right] \right\} \>_0. \nonumber \\ \label{eq:deltaA}
\end{eqnarray}
The $\<,\>_0$ in \Eq{eq:deltaA} means an average with respect to the
free fermionic action given by $A_F=\bar{\psi} D_0 \psi$, with $D_0$
connecting fermions that reside on edges of a link on
${\cal B}$, and is given by
\begin{equation}
D^{-1}_0=\frac1{\epsilon^2+q^2}\left( \begin{array}{cc} -i\epsilon &
   q \\ q & -i\epsilon \end{array} \right).
\label{eq:Do}
\end{equation}

\begin{figure}[htb]
\includegraphics[width=9cm]{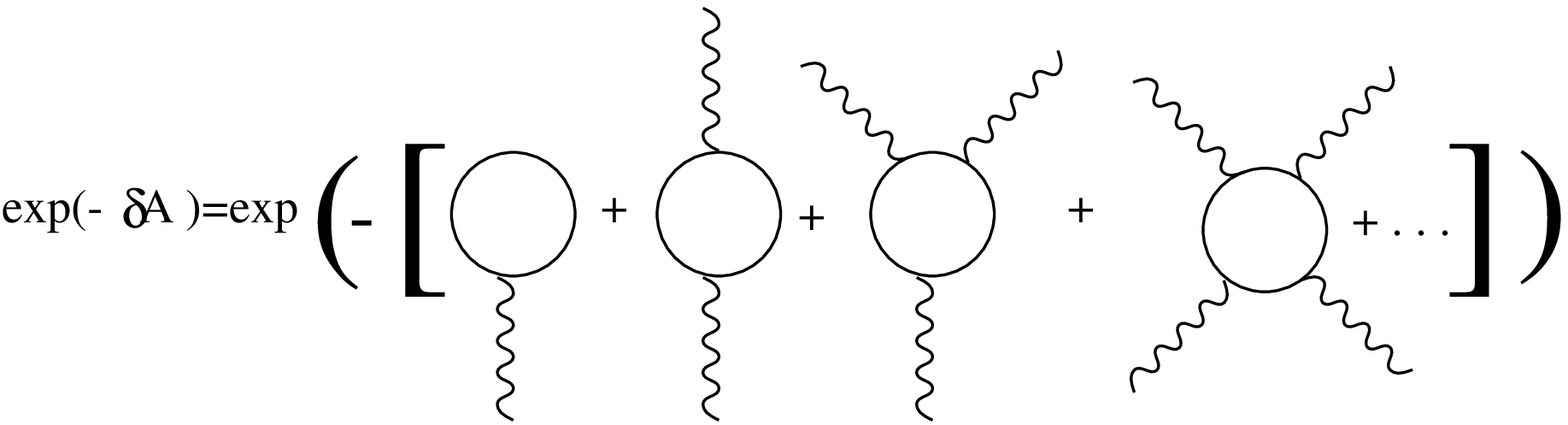}
\caption{Summing the bubble diagrams. Internal fermion loops are of
  $O(N_f)$, and each vertex with the external $Q$ and $\lambda$ fields is of
  $O(1/\sqrt{N_f})$. \label{fig:bubbles}}
\end{figure}

The calculation of $\delta A$ is straightforward and the result is
given pictorially in Fig.~\ref{fig:bubbles}, where the external legs
represent the fields $Q$ and $\lambda$.
We denote the contributions to $\delta A$ as $\sum_{n,m=1}^{4} \delta A_{Q^n \lambda^m} +
O(1/N_f^{3/2})$, according to their power in the fields $Q$ and $\lambda$, and
calculate all terms up to $n=m=4$. An important remark is that
we choose to work in a regime where $q^2\gg 1/N_f$. This is consistent
with our analysis in Section~\ref{sec:largeNf}, and in
Appendix~\ref{app:f_of_x}, and still leaves the possibility to
take $t\sim 1/\sqrt{N_f}$.
\subsection{$O(\sqrt{N_f})$ terms}
The first term in Fig.~\ref{fig:bubbles} contributes
\begin{eqnarray}
\delta A_Q&=&-\sum_{\bn\mu\in {\cal
 B}} \left( Q_{\bn\muhat}(\omega=0) + h.c.
\right) \sum_\epsilon \frac{q}{\epsilon^2+q^2} \nonumber
\frac{1}{\sqrt{4N_f\beta}} \times 4N_f \nonumber \\ &&   =
-\sqrt{4N_f\beta} \tanh(q/2T) \sum_{\bn\mu\in {\cal B}} {\text Re}\,
\,
Q_{\bn\muhat}(\omega=0), \nonumber \\ \label{eq:OsqrtN1} \\
\delta A_\lambda&=&-\sum_\bn \lambda_\bn \sum_\epsilon
\frac{1}{\beta}\left[ \frac{-i\epsilon-\frac12 \epsilon^2
    \beta/N_\tau}{\epsilon^2+q^2} + O(1/N^2_\tau) \right]  \times
\nonumber \\
&& \frac{i}{\sqrt{4N_f}} \times -4N_f \beta =
-\frac{i}{2} \beta \sqrt{4N_f} \sum_\bn \lambda_\bn,
\label{eq:OsqrtN2}
\end{eqnarray}
where $N_\tau$ is the number of Euclidean time slices. Using the MF
equation $\tanh(q/2T)=q/2J$, it is easy to show that
Eqs.~(\ref{eq:OsqrtN1}--\ref{eq:OsqrtN2}) cancel the $O(\sqrt{N_f})$
contributions in \Eq{eq:A_eff0}. This is guaranteed since
the mean-field solution is a stationary point of $A_{\text eff}$.

\subsection{$O(1)$ terms}
The second diagram in Fig.~\ref{fig:bubbles} and the
quadratic term in \Eq{eq:A_eff0}, give the mass terms of $\lambda$ and
$Q$. For $Q$ we find
\begin{eqnarray}
\delta A_{Q^2}&=&\sum_\omega \left[ \sum_{\bn \muhat\in {\cal B}}
  \left(
    \frac12 m^2 |Q_{\bn\muhat}(\omega)|^2 + v
    {\text Re}\, Q_{\bn\muhat}(\omega)Q_{\bn\muhat}(-\omega)  \right)\right.
  \nonumber \\
&+& \left. \sum_{\bn \muhat \notin {\cal B}}
    \frac12 m^2 |Q_{\bn\muhat}(\omega)|^2 + \sum_{\bn_{1,2}
    \mu \in
      {\cal B'}} 2v
    {\text Re} Q^*_{\bn_1\muhat}(\omega)Q_{\bn_2\muhat}(\omega)
  \right], \nonumber \\
\end{eqnarray}
where the links $\bn_1\muhat$, and $\bn_2\muhat$ in ${\cal B'}$ are
connected by two links in ${\cal B}$. The form of $m,v$ (see also the
related \cite{Stasio}) is
\begin{eqnarray}
\frac12 m^2&=&\frac{1}{4J}-\frac1{\beta}\sum_\epsilon
\frac{\epsilon(\omega+\epsilon)}{(\epsilon^2+q^2)((\epsilon+\omega)^2+q^2)},\\
 v&=&\frac1{\beta}\sum_\epsilon
\frac{q^2}{(\epsilon^2+q^2)((\epsilon+\omega)^2+q^2)}.
\end{eqnarray}

At $\omega=0$, and $T<T_c$ we get
\begin{equation}
m^2=2v=\frac14\left[\frac1{J}-\frac{\beta}{\cosh^2(\beta q /2)}\right],
\end{equation}
which vanishes like $T_c-T$ near $T_c=J$. In the disordered phase,
where $q=0$, then $v=0$, and $m^2(\omega=0)=(T-T_c)/(4TT_c)$. For
$\omega\neq0$, we find that close to $T_c$ (where $q=0$) the
difference $\delta m^2=m^2(\omega)-m^2(0)$ is given by
\begin{eqnarray}
\delta m^2&=&\frac{2\omega}{\beta} \sum_\epsilon
\frac1{\epsilon^2(\epsilon+\omega)}=\frac{2\omega}{\beta}
\sum_\epsilon
\frac1{(\epsilon-\omega)^2\epsilon}\nonumber \\
&&=\frac{\omega}{\beta} \left[\sum_\epsilon
\frac1{(\epsilon-\omega)^2\epsilon}-\frac1{(\epsilon+\omega)^2\epsilon}\right]\nonumber
\\
&=&\frac{4\omega^2}{\beta} \sum_\epsilon
\frac{1}{(\epsilon-\omega)^2(\epsilon+\omega)^2}>0.
\end{eqnarray}
Finally, the $\lambda$ fields are all massive
\begin{eqnarray}
\delta A_{\lambda^2}&=&\frac12 \sum_\bn \lambda^2_\bn \sum_\epsilon
\frac{\epsilon^2}{(\epsilon^2+q^2)^2}\nonumber \\
&&-\sum_{\bn\mu\in {\cal
B}}\lambda_\bn \lambda_{\bn+\muhat} \sum_{\epsilon}
\frac{q^2}{(\epsilon^2+q^2)^2}, \label{eq:L2}
\end{eqnarray}
and $\delta A_{\lambda Q}=0$.

\subsection{$O(1/\sqrt{N_f})$ terms}
The cubic terms come from the third diagram in
Fig.~\ref{fig:bubbles}. There are contributions of the type
$Q^3$,$Q^2\lambda$, $Q\lambda^2$, and $\lambda^3$. Provided that we
are interested only in the regime $q^2\sim 1/\sqrt{N_f}$, one needs
to keep track of terms which are of $O(q^2/\sqrt{N_f})$. This leaves
us with the following
\begin{eqnarray}
\delta A^-_{Q^3} &=&\frac{1}{\sqrt{4N_f}} \sum_{\bn\mu\in {\cal B}
\atop \omega_{1,2}} {\text Re} \left(
Q_{\bn\muhat}(\omega_1)Q_{\bn\muhat}(\omega_2)\nonumber \right. \\
&&\left. \cdot Q_{\bn\muhat}^*(\omega_1+\omega_2)\right) \times \left(V^{(3)}+O(q^3)\right), \\
\delta A^{--}_{Q^3} &=&\frac{1}{\sqrt{4N_f}} \sum_{\bn\mu\in {\cal
B} ,\nu \atop \omega_{1,2}} {\text Re} \left(
Q_{\bn\nuhat}(\omega_1)Q_{\bn\muhat}(\omega_2)\right. \nonumber \\
&& \left. \cdot Q_{\bn\muhat}^*(\omega_1+\omega_2)
\right) \times V^{(3)}, \\
\delta A^{\sqcap}_{Q^3} &=&\frac{1}{\sqrt{4N_f}} \sum_{\bn\nu\in
{\cal B},\mu \atop \omega_{1,2}} {\text Re} \left(
Q^*_{\bn+\nuhat,\muhat}(\omega_1)Q_{\bn\muhat}
(\omega_2)\right. \nonumber \\
&& \left. \cdot Q_{\bn+\muhat,\nuhat}(\omega_1-\omega_2) \right) \times
V^{(3)}.
\end{eqnarray}
where by ``$--$'' we mean paths along two links, with one link in
${\cal B}$, and by ``$\sqcap$'' we mean staples that begin and end
on the edges of a link in ${\cal B}$. The cubic coupling $V^{(3)}$
depends on the external frequencies, and is given by
\begin{eqnarray}
 V^{(3)}(\omega_1,\omega_2)&=& \frac2{\beta^{3/2}} \sum_{\epsilon}
\frac{q}{\epsilon^2+q^2} \times
\frac{\epsilon+\omega_2}{(\epsilon+\omega_2)^2+q^2} \nonumber \\
&&\times
\frac{\epsilon-\omega_1}{(\epsilon-\omega_1)^2+q^2}.
\end{eqnarray}
 Next we find that
$\delta A_{Q^2\lambda}$ is given by
\begin{eqnarray}
\delta A_{Q^2\lambda}&=&\frac{1}{\sqrt{4N_f}} \sum_\omega \left[
\sum_{\bn\mu} |Q_{\bn\muhat}(\omega)|^2
(\lambda_\bn+\lambda_{\bn+\muhat}) \nonumber \right. \\ && \left. \times
 \frac1{\beta^{3/2}}\sum_\epsilon
\frac{\epsilon^2(\omega-\epsilon)}{(\epsilon^2+q^2)((\omega-\epsilon)^2+q^2)}\right. \nonumber\\
&&\left.+\sum_{\bn\mu\in{\cal B}}|Q_{\bn\muhat}(\omega)|^2
(\lambda_\bn+\lambda_{\bn+\muhat}) \right. \nonumber \\ && \left. \frac1{\beta^{3/2}}\sum_\epsilon
\frac{q^2(\epsilon-\omega)}{(\epsilon^2+q^2)((\omega-\epsilon)^2+q^2)}\right].
\label{eq:Q2L}
\end{eqnarray}
Here note that both contributions of \Eq{eq:Q2L} do not involve the
zero modes, since for $\omega=0$, the sums over $\epsilon$ vanish.
Next we find
\begin{eqnarray} \delta
A_{\lambda^2 Q} &=& -\frac{1}{\sqrt{4N_f}} \sum_{\bn\mu\in {\cal B}}
{\text Re} \, Q_{\bn \muhat}(\omega=0)\left( \lambda^2_\bn
+\lambda_\bn\lambda_{\bn+\muhat}+\lambda^2_{\bn+\muhat}\right)\nonumber
\\ && \times
\frac2{\beta^{3/2}}\sum_{\epsilon} \frac{q\,
\epsilon^2}{(\epsilon^2+q^2)^3}, \label{eq:L2Q}
\end{eqnarray}
and that $\delta A_{\lambda^3}=0$.

\subsection{$O(1/N_f)$ terms}
Here we have five type of interactions terms: $\delta A_{Q^4}$,
$\delta A_{Q^3\lambda},\delta A_{Q^2\lambda^2},\delta
A_{Q\lambda^3},\delta A_{\lambda^4}$. Out of these, we present only
$\delta A_{Q^4}$. The reason is that $\delta A_{Q^3\lambda}$, and
$\delta A_{Q\lambda^3}$ are $O(q/N_f)$, and that the $O(1/N_f)$
terms, $\delta A_{\lambda^4}$ and $\delta A_{\lambda^2Q^2}$,
contribute negligible
 corrections to the effective action of the $Q$ fields (see
 Appendix~\ref{app:Intmassive}). What we find is

\begin{eqnarray}
\delta A^-_{Q^4} &=& \frac{1}{4N_f} \sum_{\bn\mu \atop
\omega_{1,2,3}} {\text Re} \left(
Q_{\bn\muhat}(\omega_1)Q^*_{\bn\muhat}(\omega_2)Q_{\bn\muhat}(\omega_3)
\right. \nonumber \\
&& \left. Q_{\bn\muhat}^*(\omega_1+\omega_3-\omega_2)
\right) \times \frac12 V^{(4)},\\
\delta A^{--}_{Q^4} &=&\frac{1}{4N_f} \sum_{\bn\mu \nu \atop
\omega_{1,2,3}} {\text Re} \left(
Q_{\bn\muhat}(\omega_1)Q^*_{\bn\muhat}
(\omega_2)Q_{\bn\nuhat}(\omega_3) \right. \nonumber \\
&& \left. Q_{\bn\nuhat}^*(\omega_1+\omega_3-\omega_2) \right) \times
V^{(4)},\\
\delta A^{P}_{Q^4} &=&\frac{1}{4N_f} \sum_{\bn\mu\nu \atop
\omega_{1,2,3}} {\text Re} \left(
Q_{\bn\muhat}(\omega_1)Q^*_{\bn+\nuhat,\muhat}(\omega_2)Q_{\bn+\muhat,\nuhat}(\omega_3)
\right. \nonumber \\
&& \left. Q_{\bn\nuhat}^*(\omega_1+\omega_3-\omega_2) \right) \times
2V^{(4)}, \label{eq:Q4P}
\end{eqnarray}
where ``$--$'' we again mean all self-avoiding paths of length two,
while the $\delta A^P$ term includes self-avoiding closed paths of
length four, i.e. plaquettes. The quartic coupling is given by
\begin{eqnarray}
 V^{(4)}(\omega_1,\omega_2,\omega_3)&=& \frac1{\beta^{2}} \sum_{\epsilon}
\frac{\epsilon}{\epsilon^2+q^2} \times
\frac{\epsilon+\omega_1}{(\epsilon+\omega_1)^2+q^2} \nonumber
\\
&&\times\frac{\epsilon+\omega_1-\omega_2}{(\epsilon+\omega_1-\omega_2)^2+q^2}\nonumber
\\ && \times
\frac{\epsilon+\omega_1+\omega_3-\omega_2}{(\epsilon+\omega_1+\omega_1-\omega_2)^2+q^2}.
\end{eqnarray}

\section{Calculation of $A^0_{\text eff}$}
\label{app:Intmassive}

In this section we integrate over the massive fields
$Q(\omega\neq0)$ and $\lambda$. (The gauge transformations are of
course still massless, but do not cause any divergences in the path
integral, since the gauge group is compact). To integrate over
$\lambda$ we can assume that it has only a mass term, as close to
$T_c$ the second term in \Eq{eq:L2} is negligible. The effective
action for the link fields $A_{\text eff}(Q)$ is defined by
\begin{eqnarray}
\exp(-A_{\text eff}(Q))&=&\exp(-\sum_n \delta A_{Q^n\lambda^0})
%\nonumber \\ &&
\times
\< e^{-\delta A_{Q\lambda^2}-\delta A_{Q^2\lambda}-\delta
A_{Q^2\lambda^2}} \>,
\end{eqnarray}
where by $\<,\>$ we mean an average with respect to $\delta
A_{\lambda^2}$, which we evaluate by expanding the exponential in
$1/N_f$. Here we have neglected $\delta A_{\lambda^4}$, which is of
$O(1/N_f)$, and will contribute to $A_{\text eff}(Q)$ only through
$\<\delta A_{\lambda^4} \delta A_{Q\lambda^2} \>$ which is of
$O(q/N^{3/2}_f)$.

Close to $T_c$, only the first term in \Eq{eq:L2} is important,
which means that only even terms in this expansion will survive.
This leaves us with the following contributions. From $\<\delta
A_{Q\lambda^2}\>$ we get a term of $O(q/\sqrt{N_f})$, which is
linear in ${\text Re} Q(\omega=0)$. Together with \Eq{eq:OsqrtN1}
and the first $O(\sqrt{N_f})$ term in \Eq{eq:A_eff0}, this yields a
$1/N_f$ correction for the MF equation, which can be absorbed in a
$O(1/N_f)$ correction to $T_c$. From $\<\delta A_{Q^2\lambda^2}\>$
we get a $1/N_f$
 contribution to the square masses of the $Q$ fields.
Since the masses of the $\omega \neq 0$ fields, and of the zero
modes are of $O(1)$, and $O(q^2)$, respectively, we neglect this
contribution as well. In fact, since the square mass of the zero
modes scales like $|T-T_c|$, this contribution can also be
considered as an $O(1/N_f)$ correction to $T_c$. Finally from
$\<(\delta A_{Q^2\lambda})^2\>\sim O(1/N_f)$ we get an $O(1/N_f)$
contribution to the quartic interactions of the {\em massive} modes,
$Q(\omega\neq0)$.

We now proceed to integrate out the fields $\tilde{Q}\equiv
Q(\omega\neq0)$. The starting point is the action
\begin{equation}
A_{\text eff}(Q)=\sum_n \delta A_{Q^n\lambda^0} - \frac12 \<(\delta
A_{Q^2\lambda})^2\>.
\end{equation}
Writing the first expression on the right hand side in terms of the
zero modes $\phi=Q(\omega=0)$, and the massive fields $\tilde{Q}$,
one finds that interactions of the type $\tilde{Q}^2\phi\sim
O(q/\sqrt{N_f})$, and $\tilde{Q}^2\phi^2,\tilde{Q}^3\phi\sim
O(1/N_f)$. Repeating the steps we took to integrate over $\lambda$,
one finds almost the same conclusions; there are $O(1/N_f)$ corrections to
the square mass of the zero modes and to the MF equation, but there
are no corrections to the quartic interactions of the zero modes.

In light of the above, the effective action of the zero modes,
$A^0_{\text eff}(\phi)$, is given by $A_{\text eff}(Q)$ when setting
$\tilde{Q}=0$. This gives \Eq{eq:A_eff_zero_T-}, with the masses
$m_{\bn\muhat}=m(\omega=0)$, and with the cubic and quartic terms,
at zero external frequencies $\omega_1=\omega_2=\omega_3=0$. For our
discussion in Section~\ref{sec:largeNf}, it is sufficient to write
the effective action for the zero modes {\em above} $T_c$. This is
simpler, since there $q=0$, which makes all the mass differences and
the cubic interactions vanish. The result is
\begin{eqnarray}
A^0_{\text eff}&=&\sum_{\bn \mu} \frac12 m^2 |\phi_{\bn \mu}|^2
%\nonumber \\ &&
+
\frac{V^{(4)}}{4N_f} \left\{ \sum_{\bn\mu} \frac12  |\phi_{\bn
\mu}|^4 + \sum_{\bn\atop \mu\neq\nu} |\phi_{\bn \mu}|^2
|\phi_{\bn+\muhat,\nu}|^2  \right. \nonumber \\ && \left.  +  2{\text Re} \left( \phi_{\bn \mu} \phi_{\bn+\muhat,\nu}
\phi^*_{\bn+\mu+\nu,\mu} \phi^*_{\bn+\nu,\mu} \right)
\right\}. \label{eq:A_eff_zero_Tp_1}
\end{eqnarray}
Here the second and third quartic interactions are between adjacent
links lattice, and links on a common plaquette. Also we have
$m^2=(T-T_c)/(4TT_c)$, and $V^{(4)}=\frac1{\beta^2}\sum_\epsilon
\epsilon^{-4}=1/(48T^2)$. The action \Eq{eq:A_eff_zero_Tp_1} has a
 manifest $U(1)$ gauge symmetry, which means that
some of the fluctuations will have an identically zero mass. This
action is also invariant under lattice translations and rotations,
which are the symmetries that the spin-Peierls state breaks. Using the
notations of Fig.~\ref{fig:fcc_lattice} we introduce the Fourier transform
\begin{equation}
\phi_{\bk i}=\sqrt{\frac2{N_s}}\sum_{\bN} \phi_{\bN i} e^{i\bN
\bk}\times
\left[ \begin{array}{cc} 1 & i=1 \\
e^{ik_y/2} & i=2 \\
e^{i(k_x+k_y)/2)} & i=3 \\
e^{ik_x/2} & i=4 \end{array}. \right. \label{eq:FT}
\end{equation}
Here $\bN$ denotes a site on the even sublattice, and $\bk$ denotes a
momentum in this lattice's first Brillouin zone. In momentum space the
action looks like
\begin{eqnarray}
A&=&\sum_{\bk i} \frac12 m^2 |\phi_{\bk i}|^2 + \frac{1}{48T_c^2
  \, 4N_f \, N_s/2}
\sum_{\{\bk\}} \left\{ \frac12 \sum_i \phi_{\bk_1 i}\phi^*_{\bk_2
i}\phi_{\bk_3 i} \phi^*_{\bk_4 i} \right. \nonumber \\
&& \left. + \sum_{i> j} V_{ij}(\bk_3-\bk_4) \phi_{\bk_1
i}\phi^*_{\bk_2 i}\phi_{\bk_3 j} \phi^*_{\bk_4 j}
% \nonumber
%\right. \\ && \left.
+V_P(\{\bk\})
{\text Re} \left( \phi_{\bk_1 1} \phi^*_{\bk_2 3} \phi_{\bk_3 2}
\phi^*_{\bk_4 4} \right)
\right\}\delta_{\bk_1-\bk_2+\bk_3-\bk_4},\nonumber \\
\label{eq:A_eff0_p}
\end{eqnarray}
where $\bk$ belongs to the Brillouin zone of the even sublattice,
 and $V_{12}=V_{34}=2\cos(\hat{y}(\bk_3-\bk_4)/2)$,
$V_{14}=V_{23}=2\cos(\hat{x}(\bk_3-\bk_4)/2)$,
$V_{13}=2\cos((\hat{x}+\hat{y})(\bk_3-\bk_4)/2)$,
$V_{24}=2\cos((\hat{x}-\hat{y})(\bk_3-\bk_4)/2)$, and $V_P=4\cos (
(\bk_2(\hat{x}-\hat{y})+\bk_3 \hat{y} +\bk_4 \hat{x})/2)$.

\section{Calculation of the scaling function $f(x)$}
\label{app:f_of_x}

\begin{figure}[htb]
\begin{center}
        \epsfig{width=7cm,file=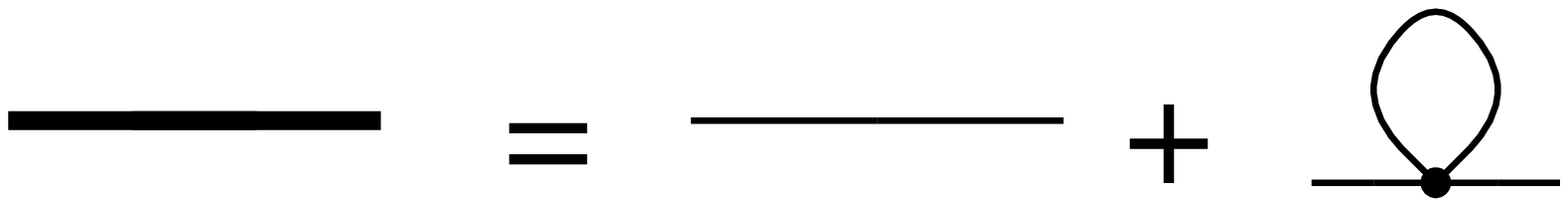} \caption{The
          calculation of $\chi$. The thin lines are
        equal to $(\frac12 m^2)^{-1}$, and the vertex can be read
        from \Eq{eq:A_eff0_p}.}\label{fig:dyson}
\end{center}
\end{figure}

To see the crossover behavior of \Eq{eq:A_eff_zero_T+} we study the
susceptibility in similar lines to the discussion in \cite{Pelissetto1,Pelissetto2}.
\begin{equation}
\chi_{ij}(\bk,\bk')= \frac{\int D\phi D\phi^* \exp (-A^0_{\text
eff}) \, \phi_{\bk i} \phi^*_{\bk' j}}{\int D\phi D\phi^* \exp
(-A^0_{\text eff})}.
\end{equation}
If $N_f=\infty$ then $A^0_{\text eff}$ contains only quadratic
terms, and $\chi\sim t^{-1}$, where $t$ is the reduced temperature
$(T-T_c)/T_c$. To see deviations from this we expand the exponential
in $1/N_f$. The result is given pictorially in Fig.~\ref{fig:dyson},
and we find that $\chi_{ij}(\bk,\bk')=\chi \, \delta_{ij}
\delta_{\bk,\bk'}$, while $\chi$ is
\begin{equation}
\chi = \frac{1}{\frac12 m^2} -  \frac{1}{\frac12 m^2} \times
\frac{\left[ \frac{1}{48 T_c^2 4N_f }  \left( \frac12 + 2 \cdot 3
\right) \right]}{\frac12 m^2} \times \frac{1}{\frac12 m^2}.
\label{eq:Dyson}
\end{equation}
Here the term in the brackets comes from the vertices
\Eq{eq:A_eff0_p}, and also counts the number of fields that run in
the loop. In terms of $t$ we find that the susceptibility is
\begin{equation}
\chi = 8T_c t^{-1} \left[ 1 - \frac{13}{6} \frac1{N_ft^2} \right].
\label{eq:chi1}
\end{equation}
Defining $x\equiv t\, \sqrt{\frac{6}{13}N_f}$, and comparing \Eq{eq:chi1} with \Eq{eq:f}, we identify $f(x)=1-1/x^2$ and $p=1/2$. More important, \Eq{eq:chi1} means
that fluctuations become significant when $x\sim O(1)$, indicating a
crossover from MF (at $x=\infty$) to a nontrivial fixed point of the
effective theory in
Eqs.~(\ref{eq:A_eff_zero_T-}),(\ref{eq:A_eff_zero_T+}). This occurs
at $(T-T_c)/T_c \sim 1/\sqrt{N_f}$.

\begin{acknowledgments}
I thank F.~Bursa, A.~Pelissetto, M.~Teper, and M.~Veillette for useful
discussions, P.~Calabrese for pointing my attention to reference
\cite{Pelissetto1}, and B.~Svetitsky for his helpful remarks on this manuscript.
I was supported by PPARC.
\end{acknowledgments}

\end{document}